\documentclass[conference]{IEEEtran}
\IEEEoverridecommandlockouts
\usepackage{cite}
\usepackage{amsmath,amssymb,amsfonts}
\usepackage{algorithmic}
\usepackage{graphicx}
\usepackage{textcomp}
\usepackage{xcolor}
\newcommand{\mypara}[1]{{\smallskip \noindent \bf #1}\hspace{0.1in}}

\def\BibTeX{{\rm B\kern-.05em{\sc i\kern-.025em b}\kern-.08em
    T\kern-.1667em\lower.7ex\hbox{E}\kern-.125emX}}

\ifodd 1
\newcommand{\congr}[1]{{\color{blue}#1}}
\else 
\newcommand{\congr}[1]{#}
\fi

\ifodd 1
\newcommand{\congc}[1]{{\color{red}(Cong: #1)}}
\else
\newcommand{\congc}[1]{}
\fi

\ifodd 1
\newcommand{\kunr}[1]{{\color{cyan}#1}}
\else
\newcommand{\kunr}[1]{}
\fi

\begin{document}

\title{Advancing RAN Slicing with\\ Offline Reinforcement Learning%: Strategies, Adaptability, and Insights in Resource Management
\thanks{The work is partially support by the US National Science Foundation under awards CNS-2002902, ECCS-2029978, SII-2132700, CNS-2003131, ECCS-2030026, ECCS-2143559, and Intel Corp.}
}

\author{\IEEEauthorblockN{Kun Yang$^*$, Shu-ping Yeh$^\ddag$, Menglei Zhang$^\ddag$, Jerry Sydir$^\ddag$, Jing Yang$^\dag$, and Cong Shen$^*$}
\IEEEauthorblockA{$^*$ Department of Electrical and Computer Engineering, University of Virginia, USA\\
$^\dag$ School of Electrical Engineering and Computer Science, The Pennsylvania State University, USA\\
$^\ddag$ Intel Corporation, USA
}}

\maketitle

\begin{abstract}
Dynamic radio resource management (RRM) in wireless networks presents significant challenges, particularly in the context of Radio Access Network (RAN) slicing. This technology, crucial for catering to varying user requirements, often grapples with complex optimization scenarios. Existing Reinforcement Learning (RL) approaches, while achieving good performance in RAN slicing, typically rely on online algorithms or behavior cloning. These methods necessitate either continuous environmental interactions or access to high-quality datasets, hindering their practical deployment. Towards addressing these limitations, this paper introduces offline RL to solving the RAN slicing problem, marking a significant shift towards more feasible and adaptive RRM methods. We demonstrate how offline RL can effectively learn near-optimal policies from sub-optimal datasets, a notable advancement over existing practices. Our research highlights the inherent flexibility of offline RL, showcasing its ability to adjust policy criteria without the need for additional environmental interactions. Furthermore, we present empirical evidence of the efficacy of offline RL in adapting to various service-level requirements, illustrating its potential in diverse RAN slicing scenarios.

% Dynamic radio resource management (RRM) in the ever-expanding and diverse landscape of wireless networks presents significant challenges, particularly in the context of Radio Access Network (RAN) slicing. This technology, crucial for catering to varying user requirements, often grapples with complex optimization scenarios. Traditional Reinforcement Learning (RL) approaches, while effective in RAN slicing applications, typically rely on online algorithms or behavior cloning. These methods necessitate either continuous environmental interactions or access to high-quality datasets, hindering their practical deployment. Towards addressing these limitations, this paper introduces offline RL to solving the RAN slicing problem, marking a significant shift towards more feasible and adaptive RRM methods. We demonstrate how offline RL can effectively learn (near)-optimal policies from sub-optimal datasets, a notable advancement over existing practices. Our research highlights the inherent flexibility of offline RL, showcasing its ability to adjust policy criteria without the need for additional environmental interactions. Furthermore, we present empirical evidence of the efficacy of offline RL in adapting to various service-level requirements, illustrating its potential in diverse RAN slicing scenarios.
\end{abstract}


\section{Introduction}
\label{sec:intro}

% Context and Motivation:.

In the rapidly evolving landscape of wireless communication, Radio Access Network (RAN) slicing plays an important role in providing heterogeneous services to diverse wireless network users, offering a paradigm shift towards more flexible and efficient use of the network resources. At its core, RAN slicing involves partitioning a single physical network into multiple virtual networks, each tailored to meet a set of {specific service requirements}. This flexibility is pivotal in addressing the diverse demands of modern wireless communications, ranging from high-speed data services to massive machine-type communications. By enabling dynamic allocation and optimization of network resources, RAN slicing significantly enhances network efficiency, scalability, and service customization. It is a key technology in the evolution of wireless networks, facilitating the transition to more adaptive, service-oriented architectures \cite{foukas2017network, afolabi2018network}. 

% The significance of RAN slicing lies not only in its ability to cater to varied service demands but also in its potential to unlock new business models and opportunities in the telecommunications sector.

However, the implementation of RAN slicing introduces complex challenges, particularly in Radio Resource Management (RRM) \cite{guo2019enabling, akhtar2021radio}. The need for customized and sophisticated RRM strategies becomes paramount to ensure near-optimal performance of each network slice without compromising the overall network integrity. Recently, Reinforcement Learning (RL) emerges as a promising tool \cite{meng2019power,ahmed2019deep,naderializadeh2021resource,yang2023mixture}, offering adaptive and intelligent solutions to navigate the intricate RRM landscape. The ability of RL to learn and make decisions based on dynamic network environments makes it ideally suited for managing the unique demand of each network slice. 

% Problem Statement:.

In this direction, RRM for RAN slicing has predominantly utilized \textbf{online} RL algorithms\cite{zangooei2023reinforcement,azimi2022applications}. While these methods excel in offering real-time adaptability, they are not without drawbacks. First and foremost, online RL algorithms require continuous interactions with the environment, a process that can be resource-intensive and potentially harmful to the performance of a wireless system, given the exploratory nature of online RL. Such extensive environmental engagement can prove costly and disruptive in real-world applications. Secondly, these online solutions are often highly specialized, designed for specific network conditions. When faced with environmental variations, they tend to exhibit sub-optimal performances, necessitating extensive retraining or fine-tuning to adapt to each different environment. This inflexibility presents considerable challenges when deploying these solutions across diverse slices, each with its own set of service level agreements (SLAs). Lastly, adapting these algorithms to even slight changes in target objectives or network parameters -- such as shifting from a throughput-sensitive to a latency-sensitive policy -- requires further training. This adaptation process inevitably leads to more environmental interactions. The limitations of current online RL approaches in RRM for RAN slicing highlight an urgent need for more robust, versatile, and efficient RL strategies. 

As an alternative, \textbf{offline} RL \cite{levine2020offline} offers a solution that is less interaction-intensive and more adaptively robust. Compared with online RL, offline RL significantly reduces the need for environmental interaction, a key drawback of online methods, thereby lowering operational costs and minimizing potential disruptions in the live wireless system. Additionally, the adaptability of offline RL for fluctuating network conditions and objectives is much better than that of online RL, enhancing its practicality in diverse and evolving network scenarios. Due to these attractive properties, offline RL solutions have been investigated \cite{yang2023mixture}.   However, these solutions have mostly focused on behavior cloning (BC), where the performance of the trained policy cannot surpass the performance of the behavior policy (BP) used to collect the dataset. Furthermore, the trained policy can only mimic what the BP does, showing little flexibility during potential SLA adaptation. Offline RL, on the other hand, has much better adaptivity compare to both online solution and previously existing BC solutions. 

% Contribution: 

In this paper, we introduce  offline RL to solving RRM problems in RAN slicing, exploring its potential in addressing complex network management challenges. Our research focuses on the adaptability of offline RL when trained on diversely collected datasets. The contributions of this study are three-fold, each highlighting a distinct aspect of offline RL in the context of RRM in RAN slicing.

\mypara{Learning from Sub-optimal Data:} We present findings indicating that offline RL has the potential to obtain near-optimal policies even when trained on sub-optimal datasets. This suggests that offline RL might be less dependent on the quality of data compared to traditional methods, a promising direction for scenarios where optimal data is not available.

\mypara{Adapting to Different SLA Requirements:} Our study explores how offline RL can potentially adapt to different SLA requirements. The result indicates that, by training across datasets with varying SLA conditions, offline RL could adjust its strategies to meet the specific needs of different network slices, a valuable feature for managing diverse network demands.

\mypara{Behavioral Flexibility with Tailored Reward Functions:} We also investigate the ability of offline RL to alter its behavior by changing the reward functions during offline training, even when using the same dataset. This result suggests that offline RL could offer a flexible approach to RRM, adapting to various operational objectives without the need for additional data.

These contributions aim to shed light on the potential adoption of offline RL in RRM for RAN slicing. While preliminary, our findings suggest that offline RL could be a valuable tool in developing more adaptable and efficient wireless network RRM solutions, capable of responding to the diverse and ever-changing demands of modern communication systems. This contributes to the ongoing discourse on the application of RL in complex network environments and sets the stage for further exploration in the field.

Beside the aforementioned contributions, we also highlight a unique feature of our work. Instead of using a white-box simulator as the environment, where the researchers have full control of all the environmental configurations that can lead to ``cheap'' data collection and environment interaction. All results in this work are obtained by utilizing a black box environment that was developed by Intel Labs, where the learning agent has very limited access to the simulated environment and interaction is very costly (mainly from the time perspective). 

% can only configure a few aspects of the environment and need to pay more expense (mainly time wise) to interact with the environment, which is more practical compare to the previous works.   

% Paper Structure: Briefly outline the structure of your paper.

% The structure of the remainder of our paper is organized as follows: Section \ref{sec:related} provides a comprehensive review of the existing literature, covering efforts to enhance online RL performance for RAN slicing as well as existing offline RL methodologies. In Section \ref{sec:prob}, we delve into the formulation of the resource management problem in our wireless system, elucidating how we translate this into a Reinforcement Learning (RL) framework. Following this, Section \ref{sec:exp} details our experimental approach, highlighting how offline RL was applied and elucidating the mechanisms enabling SLA and objective adaptation within this context. Section \ref{sec:dis} is dedicated to discussing key insights and observations from applying offline RL to RAN slicing, offering a critical thinking of our findings. Finally, we conclude the paper by summarizing our work, reflecting on its implications, and suggesting potential future research directions in this evolving field.

The remainder of this paper is organized as follows. Section \ref{sec:related} provides a comprehensive review of the existing literature. The formulation of the RRM problem for network slicing and how it can be posed in an RL framework are discussed in Section \ref{sec:prob}. Section \ref{sec:exp} details our experimental settings, highlighting how offline RL is applied and elucidating the mechanisms enabling SLA and objective adaptation within this context. Key insights and observations from the experiments are discussed in Section \ref{sec:dis}. Finally, Section \ref{sec:con} concludes the paper. 

\section{Related Works}
\label{sec:related}

\mypara{RL for RAN Slicing:} RL has emerged as a popular solution for RRM in RAN slicing, with recent studies primarily focusing on enhancing RL methods to improve the wireless system performance. Novel solutions include integrating Generative Adversarial Networks (GANs) \cite{hua2019gan, kasgari2020experienced} and Long Short-Term Memory (LSTM) networks \cite{li2020lstm} for better learning, as well as exploring multi-agent systems \cite{hu2022inter, zhou2021ran} to improve system scalability. However, these online RL approaches often overlook the costs associated with continuous real-world environment interaction. In contrast, offline RL methods, which rely on behavior cloning \cite{nagib2021transfer, liu2021onslicing, polese2022colo}, face the challenge of acquiring high-quality datasets, leading to a chicken-and-egg dilemma. Our research aims to bridge these gaps, showcasing the potential of offline RL in efficiently managing RAN slicing without the constraints of constant environmental interactions or dependency on high-quality expert data.

\mypara{Offline RL:} Offline RL represents a significant departure from traditional online RL, primarily by learning policies exclusively from offline datasets. It is a focused area that has attracted substantial interest in recent RL research; see \cite{levine2020offline} for a recent tutorial. A key challenge for offline RL is its inability to directly update policies through environmental interaction, necessitating innovative approaches to overcome potential distributional shifts. To this end, several state-of-the-art model-free deep offline RL algorithms have been developed. Notable among these are batch-constrained Q-learning (BCQ) \cite{fujimoto2019off}, conservative Q-learning (CQL) \cite{kumar2020conservative}, implicit Q-learning (IQL) \cite{kostrikov2021offline}, Twin Delayed DDPG with Behavior Cloning (TD3-BC)\cite{chu2020motion}, Decision Transformer \cite{zheng2022online}, and Soft Actor-Critic for Networks (SAC-N)\cite{an2021uncertainty}. These algorithms are recognized for their effectiveness in the offline RL landscape. Furthermore, there is a growing emphasis on the theoretical aspects of optimal offline RL, with research delving into data coverage \cite{rashidinejad2021bridging,xie2021policy} and the importance of critical states \cite{kumar2022should}. Due to these attractive features, there are recent efforts in applying (or developing tailored) offline RL solutions to wireless RRM problems \cite{yang2023mixture}. We note, however, that while \cite{yang2023mixture} has explored the use of offline RL in RRM, their work does not fully address the flexibility in adapting to varying SLAs and changing policy objectives in wireless systems.

\section{Problem Setting}
\label{sec:prob}

We begin with an in-depth exploration of the wireless RRM problem that we aim to address. We emphasize the importance of flexibility in adapting to different SLA requirements and optimization objectives, illustrating why this adaptability is crucial for efficient RAN slicing systems. To facilitate a comprehensive understanding, we start with the context of RAN slicing in Section \ref{subsec:slicing_env}, setting the stage for our analysis. Building upon this foundation, we then methodically illustrate how this RRM challenge can be aptly formulated in an RL problem in Section \ref{subsec:rl}. This formulation is pivotal as it lays the groundwork for applying advanced RL techniques, including offline RL methods, to effectively manage and optimize resource allocation in RAN slicing.

\subsection{RRM for RAN Slicing}
\label{subsec:slicing_env}

\subsubsection{RRM as An Optimization Problem}

\begin{figure}[hpbt]
    \centering
    \includegraphics[width = .8\columnwidth]{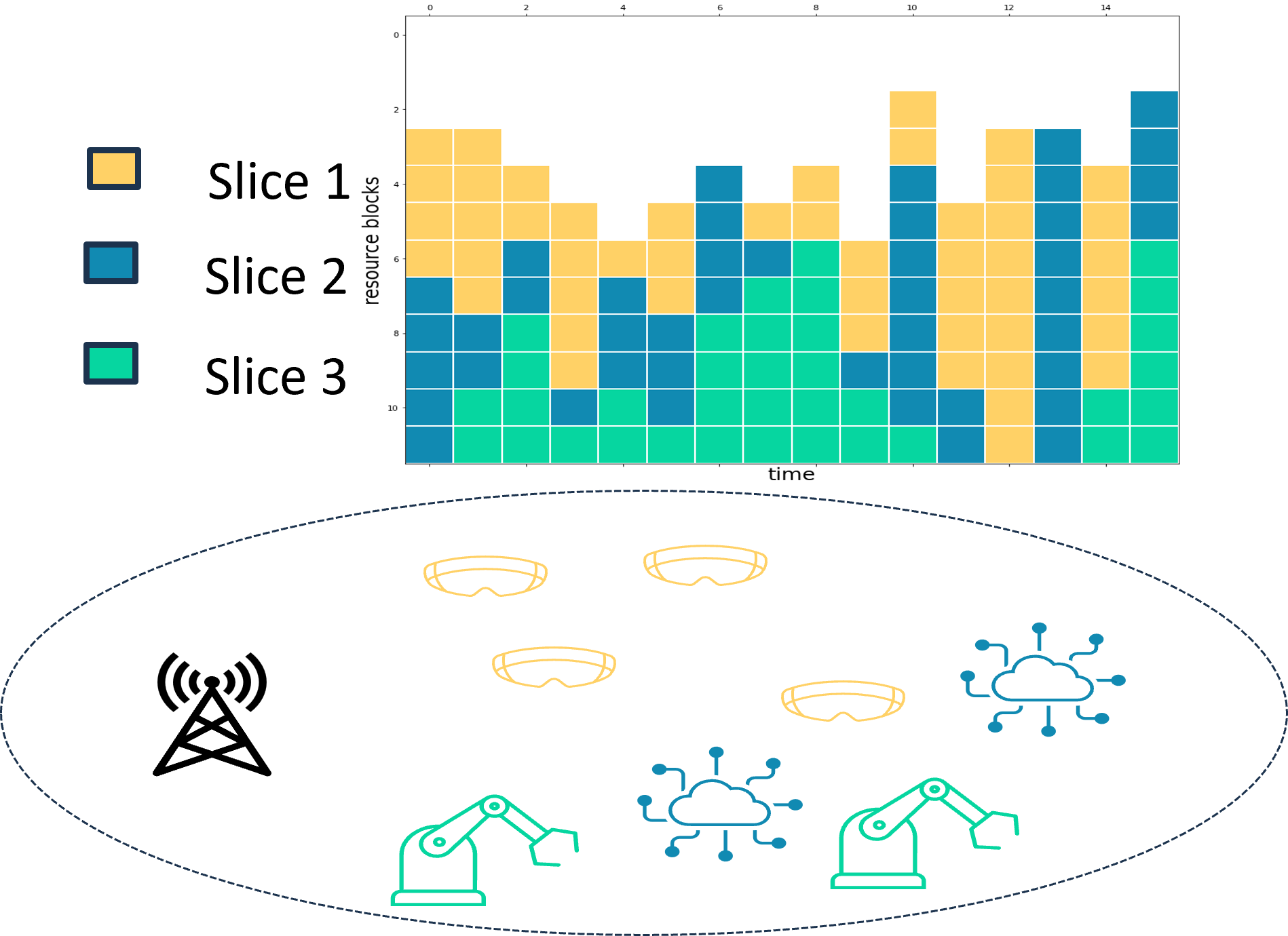}
    \caption{An illustration of the RRM system in RAN slicing, where different resource blocks are allocated to different slices with distinctive purposes.}
    \label{fig:slicing_sys}
\end{figure}

We focus on a scenario within our system that involves one cell with $N$  slices. Here, the first $N-1$ slices are designated as high-priority, while the final slice is allocated for background traffic. Each slice comprises a set of users, denoted as ${k_1, k_2, \cdots, k_N}$. Our analysis unfolds in discrete time slots, labeled as $t$, during which radio resources need to be allocated to the first $N - 1$ slices. These resources are organized into block groups, with a total of $M$ resource block groups (RBGs) available. The considered RRM system is illustrated in Figure \ref{fig:slicing_sys}.

The primary objective of this RRM problem is the strategic allocation of these resources, represented by $\mathcal{M} = \{ m_1, m_2, \cdots, m_{N-1} \}$, across the $N - 1$ high-priority slices. In our particular implementation, following this allocation, a scheduler within the NS-3 framework takes over to further distribute the resources among the users in each slice. 

The overarching goal of resource management is to strike an optimal balance in allocation, thereby fulfilling diverse Quality of Service (QoS) requirements captured by a utility function $f_t$. To formally represent this optimization challenge, we can articulate it as follows:

\begin{equation} 
\label{eqn:rrmopt}
\begin{aligned}
& \underset{{ m_1, m_2, \cdots , m_{N-1}}}{\text{maximize}}
& & f_t(m_1, m_2, ... , m_{N-1}) \\
& \text{subject to}
& & \sum_{i = 1}^{N-1} m_i \leq M. \
\end{aligned}
\end{equation}

The optimization formulation \eqref{eqn:rrmopt} encapsulates the essence of the RRM problem in our system. It aims to maximize the QoS function $f_t$, considering the constraints on the total available resources. The formulation underscores the need to judiciously allocate resources across slices, ensuring that high-priority slices receive the necessary resources while also catering to the background traffic needs. 

% \congc{I don't quite understand the optimization problem \eqref{eqn:rrmopt}. The objective function has nothing to do with $m_N$, which only appears in the constraint. So, isn't it always best to assign $m_N=0$ in order to maximize $f_t$?} \kunr{Yes, you are right professor, I have removed $N$ here}

\subsubsection{Choices of Resource Allocation}

We now discuss the resource allocation strategies for the RAN slicing system. We categorize resources into two primary types: dedicated and prioritized \cite{dighriri2018resource, ko2021priority}. Dedicated resources are exclusively reserved for a specific slice and cannot be utilized by others. In contrast, prioritized resources, while initially allocated to a particular slice, may be used by other slices if any residual capacity remains.  Based on these two distinction resource types, we draw our first two resource allocation strategies from the IETF report\cite{Geng2017NetworkSlicing}. Besides these two strategies, we introduce a third strategy, derived from the capabilities of the netgymenv simulator \cite{zhang2023netgym}, all three strategies are named as the following:

\begin{enumerate}
    \item \textbf{Hard Slicing \cite{Geng2017NetworkSlicing}:} This strategy involves allocating only dedicated RBGs to each slice. While it simplifies the system implementation, it can also lead to potentially inefficient resource utilization due to its rigid allocation.
    \item \textbf{Limited Soft Slicing:} This approach utilizes only prioritized RBGs. It aims for more efficient resource usage by allowing the possibility of shared resources among slices, depending on the availability.
    \item \textbf{Soft Slicing \cite{Geng2017NetworkSlicing}:} A hybrid strategy that combines both dedicated and prioritized resources, offering a balance between resource efficiency and allocation specificity.
\end{enumerate}

In this work, we recognize that while hard slicing offers simplicity and ease of implementation, it may not optimally utilize the available resources. On the other hand, soft slicing, though potentially more efficient, poses challenges in the practical deployment due to its higher complexity. Given these considerations, we opt to focus on limited soft slicing as the primary approach. This choice is motivated by the aim to achieve a more resource-efficient allocation while maintaining a feasible level of system complexity.

% \begin{figure}[hpbt]
%     \centering
%     \includegraphics{fig1.png}
%     \caption{Potentially a figure explaining different resource allocation types, not necessary here}
%     \label{fig:resource}
% \end{figure}

\subsection{Reinforcement Learning Formulation}
\label{subsec:rl}

As we have previously highlighted in Section \ref{sec:intro}, RL has become a pivotal tool for addressing RRM challenges in RAN slicing. The core rationale for employing RL in RRM lies in its adeptness at navigating the dynamic decision-making processes, which is typical in resource management. Particularly in the context of RAN slicing, as detailed in Section \ref{subsec:slicing_env}, the focus is on the sequential allocation of packed RBGs to optimize the QoS performance. The iterative learning and policy refinement capabilities of RL enable an agent to progressively navigate this complex decision space, ultimately leading to strategies that can significantly enhance the resource utilization and overall network efficacy. This successful application of RL in RRM hinges on effectively formulating the problem as a Markov Decision Process (MDP).

\begin{figure}[hpbt]
    \centering
    \includegraphics[width = .65\columnwidth]{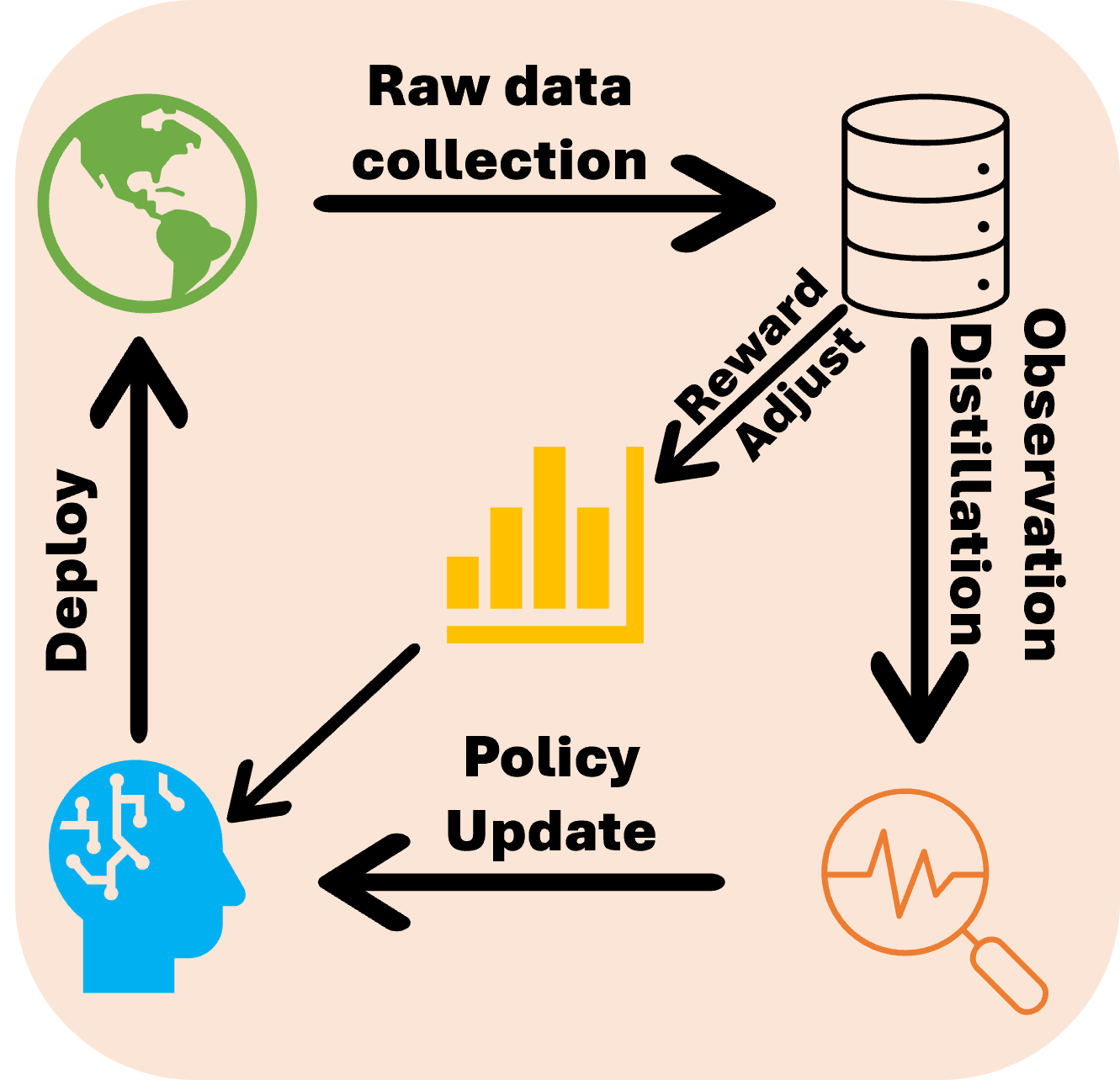}
    \caption{Offline RL with reward adjustment and observation distillation.}
    \label{fig:diagram}
\end{figure}

Diverging from the conventional methodologies that often adhere to a predetermined structure in formulating the RRM problem as an MDP, our approach introduces an innovative and more flexible paradigm as shown in Figure \ref{fig:diagram}. We break away from the standard practice of fixed observations, actions, and reward structures, instead adopting an adaptive process that is more reflective of real-world scenarios. Our system closely replicates a practical wireless network environment by capturing a comprehensive range of traffic monitoring metrics during data collection. This extensive dataset then undergoes a meticulous process of observation distillation and reward adjustment, tailored to extracting the most relevant information for the specified design objective. This nuanced approach not only brings a higher degree of realism to our simulator but also provides the adaptability necessary for a more customized RL formulation. This flexibility is critical, as it allows our system to adjust to various RRM scenarios, offering solutions that are more aligned with specific challenges and objectives encountered in real-world RAN slicing scenarios. In the following, we will elaborate on our distinct RL formulation, which we assert is well suited for addressing the intricacies of the RRM problem. %This novel approach, we believe, significantly contributes to the field by offering a more versatile and realistic framework for applying RL in RRM.

\begin{itemize}
    \item \textbf{States:} The wireless system will collect all possibly useful information from the environment, including user-level traffic load, user-level throughput, average one-way delay, maximum one-way delay, delay violation rate (with a designed threshold), resource block usage rate, and the relative location of the user. Among all the possibly useful states, we choose to collect slice-level information of throughput $T_{\text{rx}}$, traffic load $T_{\text{tx}}$, resource utilization rate $U$, delay violation rate $D_{\text{vio}}$, and average one-way delay $D_{\text{avg}}$ from every slice in the system. The states can thus be specified as:
    % \begin{equation*}
    %     \label{equ:state}
    %     [T_{\text{rx}}, T_{\text{tx}}, U, D_{\text{vio}}, D_{\text{avg}}] \times N
    % \end{equation*}
    \begin{equation*}
        \label{equ:state}
        \left\{ T_{\text{rx},i}, T_{\text{tx},i}, U_i, D_{\text{vio},i}, D_{\text{avg},i} \right \}_{i=1,\cdots, N} 
    \end{equation*}
    % \begin{equation*}
    % S(t) = 
    % \begin{bmatrix}
    % T_{\text{rx},1}(t) & T_{\text{tx},1}(t) & U_1 & D_{\text{vio},1}(t) & D_{\text{avg},1}(t) \\
    % T_{\text{rx},2}(t) & T_{\text{tx},2}(t) & U_2 & D_{\text{vio},2}(t) & D_{\text{avg},2}(t) \\
    % \vdots & \vdots & \vdots & \vdots & \vdots \\
    % T_{\text{rx},N}(t) & T_{\text{tx},N}(t) & U_N & D_{\text{vio},N}(t) & D_{\text{avg},N}(t)
    % \end{bmatrix}
    % \end{equation*}

    \item \textbf{Actions:} As stated in Section \ref{subsec:slicing_env}, our goal is to allocate the RBGs to prioritized slices, and we choose to use the limited soft slicing technique so that we are allocating prioritized resources to prioritized slices, i.e. $A(t) = [a_{1}(t), \cdots, a_{N-1}(t)]$ where $a_i(t) \in [0,1]$.

    \item \textbf{Reward:} The reward design of a RAN-slicing system should align with its QoS or the SLA. In our  setting, we care about three components: the overall throughput of the system, the delay violation rates, and the resource utilization rate. We thus design a prioritize SLA-aware reward as the following. We first define a {\it priority vector} as $\mathbf{p} = [p_1, \cdots, p_i, \cdots, p_N]$. Then the reward is given as $$ R(t) = \sum_{i=1}^{N} p_i r_i(t),$$ where $$r_i(t) = T_{\text{rx}, i}(t) - \alpha  D_{\text{vio},i}(t) - \delta  U_{i}(t).$$ 
    % \begin{equation}
    % \label{equ: reward}
    % \begin{aligned}
    % & Define \: priority \: vector: \mathbf{p} = [p_1, ... p_i, .... p_N] \\
    % & Then \: reward: R(t) = \sum_{i=1}^{N} p_i \times r_i(t) \\
    % & Where: r_i(t) = T_{\text{rx}, i}(t) - \alpha * D_{\text{vio},i}(t) - \delta * U_{i}(t)
    % \end{aligned}
    % \end{equation}
    In our experiment, we initially set $\mathbf{p} = [\frac{1}{N-1}, ..., \frac{1}{N-1}, ...., 0]$, $\alpha = 4$, and $\delta = 1$.
\end{itemize}

We note that the reward design incorporates a flexible mechanism for adjusting the priority of different slices via the priority vector 
$\mathbf{p}$. Additionally, it allows fine-tuning of the significance of each key component -- throughput, delay violation, and resource utilization -- using hyperparameters $\alpha$ and $\delta$. This flexibility is crucial for customizing the behavior of an RL agent to match the specific network conditions and SLA requirements. By varying these parameters, the reward function can be tailored to emphasizing different aspects of the network performance, thus ensuring that the learning process of an RL agent is aligned with the overarching goals of the RAN slicing system. This carefully crafted design enables the RL model to adaptively balance between maximizing throughput, minimizing delay violations, and optimizing resource usage, in accordance with the defined priorities and operational constraints.

Based on the MDP design, the objective of the RL system is given as:

$$
\underset{{\pi}}{\text{max}} \: \mathbb{E} \left[\sum_{t=0}^{\infty}\gamma^tR(t) \right].
$$

\section{Experimental Setup and Results}
\label{sec:exp}

\subsection{Simulation Setup and Baseline Methods}% in RAN Slicing Environment}
\label{subsec:setup}

We now detail the experimental setup and the initial strategies employed for data collection and performance evaluation. Our experiments are conducted using the \textbf{\texttt{netgymenv} simulator} developed by Intel \cite{zhang2023netgym}, as mentioned in Section \ref{sec:prob}. This simulation environment focuses on a RAN slicing system comprising one cell and $N$ slices. The traffic module follows the LTE module in NS-3 \cite{riley2010ns}. Specifically, our experimental setup involves a system with three slices: the first two slices are prioritized for essential services, while the third slice handles background traffic. The key parameters of our simulation are summarized in Table \ref{tab:exp}.

\begin{table}%[hpbt]
\caption{Experiment parameters}
\label{tab:exp}
\centering
\begin{tabular}{c|c}
\hline
Parameter & Value \\
\hline
Number of Slices & 3 \\
Number of UEs & $6-20$ \\
Delay violation threshold & [100, 50, 10] ms\\
Area & $120 \times 10$ m$^2$ \\
Downlink traffic & 2 Mbp/s\\
Traffic pattern & Poisson arrival\\
UE mobility & $1-2$ m/s\\
\hline
\end{tabular}
\end{table}

For delay violation rate data collection, the simulator does not record individual packet delays due to the high computational overhead. Instead, it maintains a histogram of packet arrivals over a specific period of time, from which the delay violation rate is calculated.

% Regarding data collection for offline RL training, it is initiated using behavior policies (BPs), which also serve as baselines for comparative evaluation. In our experiments, we employed the following BPs:

Regarding data collection for offline RL training, we deploy the following behavior policies (BPs), which also serve as baselines for comparative evaluation.

\begin{enumerate}
    \item \textbf{Traffic load-based resource allocation:}  Resources are allocated proportionally based on the traffic load observed in the previous time period:
    $$
    a_i(t) = \frac{T_{\text{tx},i}(t-1)}{\sum_{j} T_{\text{tx},j}(t-1) + \Delta}, a_i(0) = \frac{1}{N},
    $$
    where $\Delta$ is a small positive number to avoid numerical instability. 
    \item \textbf{Delay violation rate-based resource allocation:} Here, resource allocation is proportionally based on the delay violation rates, using a softmax function for more stable allocation:
    $$
    a_i(t) = \frac{\exp{D_{\text{vio},i}(t-1)}}{\sum_{j} \exp{D_{\text{vio},j}(t-1)}}, a_i(0) = \frac{1}{N}
    $$
    \item \textbf{Online RL:} We also include an online RL policy as a BP for both `expert' dataset collection and performance benchmarking. The specific algorithm we choose to use is Soft Actor-Critic (SAC) \cite{haarnoja2018soft}, selected for its actor-critic structure that aligns with the offline RL algorithms we deploy.
\end{enumerate}

The formulation of the first two baseline methods, Traffic Load-Based and Delay Violation Rate-Based Resource Allocation, is specifically designed to incorporate limited aspects of system information. As a result, while they are adept at optimizing certain traffic patterns, their performances may be sub-optimal when the broader system dynamics are taken into consideration. These methods, therefore, serve as useful starting points but may not fully capture the complexity and variability of real-world traffic scenarios.

In contrast, the online RL policy, specifically the SAC algorithm, is employed with the intention of reaching a more comprehensive system optimization. To ensure its effectiveness, we commit to training this policy over an extended period, allowing it to adapt and learn from a wide range of network conditions and scenarios. This extended training is crucial for the online RL policy to develop a detailed understanding of the system and achieve near-optimal performance, potentially exceeding the more narrowly focused baseline methods. The comparison between these baseline strategies and the more dynamically trained online RL policy will provide valuable insights into the effectiveness and adaptability of different resource allocation approaches in RAN slicing.

\mypara{Data Collection for Offline Datasets:} To construct comprehensive offline datasets, we systematically execute each BP for 40 episodes. An individual episode comprises 200 continuous steps. At the beginning of each episode, we introduce variability by randomly selecting both the number of users and the random seed for the prioritized slices, while maintaining a constant user count of 5 for the background slice. This approach ensures a diverse dataset that encapsulates a range of possible network states and user behaviors. By adopting this method for each BP under varying SLA requirements, we accumulate a substantial dataset of approximately 80,000 data samples per BP under the same SLA requirement. This extensive collection forms the foundational dataset for training our offline RL algorithms, providing them with a broad spectrum of scenarios to learn from and adapt to. The diversity and volume of this dataset are critical for enabling the offline RL models to develop robust and effective resource allocation strategies that can cater to the dynamic and complex needs of RAN slicing.

\subsection{Offline RL with Sub-optimal Datasets}
\label{subsec:subopt}

The initial phase of the offline RL experimentation employs Conservative Q-learning (CQL) \cite{kumar2020conservative}, a method widely known for its straightforward implementation and the pessimistic approach to offline RL training, characterized by the inclusion of a regularizer in its loss function. The specific design of CQL makes it an ideal algorithm to use as a starting point for offline RL experiments, particularly for evaluating the efficacy of offline RL in deriving practical algorithms. The training details for the online and offline RL clients are given in Table \ref{tab:train}.

\begin{table}%[hpbt]
\caption{Training parameters}
\label{tab:train}
\centering
\begin{tabular}{c|c}
\hline
Parameter & Value \\
\hline
Actor structure & 2 Layer MLP with hidden dimension 64\\
Critic structure & 2 Layer MLP with hidden dimension 64\\
Critic learning rate & $1e^{-3}$ for online, $3e^{-4}$ for offline\\
Actor learning rate & $3e^{-4}$ for online, $5e^{-5}$ for offline \\
\hline
\end{tabular}
\end{table}

Our experimental setup in this section is tailored to a scenario where both prioritized slices have identical SLA levels, though the number of users in each slice can vary. To maintain a level playing field in our comparative analysis, we standardize the definition of a training step across both online and offline RL. For online learning, a step involves collecting a data sample from the environment followed by a mini-batch gradient descent update. In the offline RL context, a step is defined as sampling a mini-batch from the dataset and performing a corresponding mini-batch gradient descent update.

\begin{figure}
    \centering
    \includegraphics[width = .8\columnwidth]{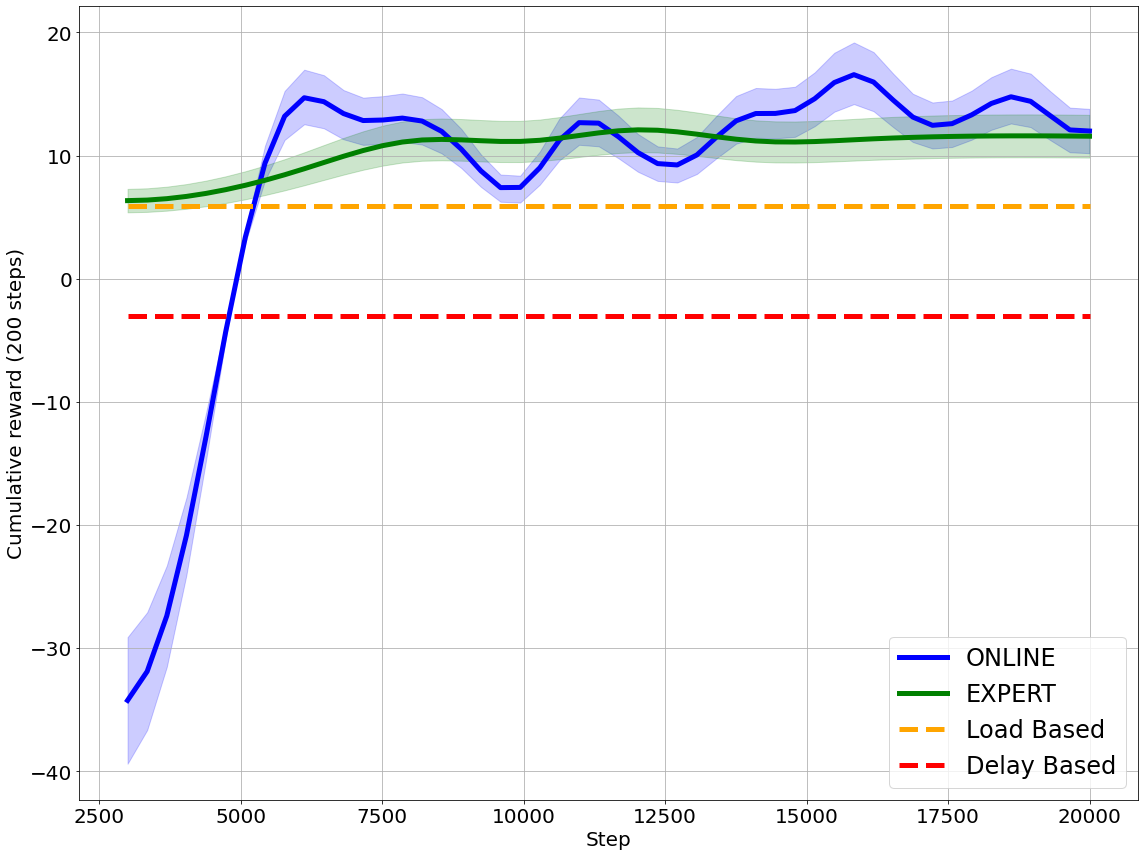}
    \caption{Offline training result with an expert dataset.}
    \label{fig:expert}
\end{figure}

As depicted in Figure \ref{fig:expert}, we present the performance results of the offline RL model, which utilizes an expert dataset collected via a SAC agent. Intriguingly, the results indicate that offline RL, when trained on a dataset acquired from a high-performing online RL agent, can surpass the performance of its online counterpart within the same number of training steps. While this outcome is noteworthy, it is important to remember that access to an expert-level dataset is not always feasible in practical systems. Therefore, a more critical assessment of the capabilities of offline RL lies in its ability to learn effective policies from \textit{sub-optimal} datasets.

In our exploration of offline RL with sub-optimal datasets, we conduct tests using the two baseline strategies outlined in Section \ref{subsec:setup}. Initially, we train an offline RL agent separately on datasets generated from each of these baselines. Subsequently, we combine these two datasets to assess any potential performance benefits from this mixed dataset approach.

\begin{figure}
    \centering
    \includegraphics[width = .8\columnwidth]{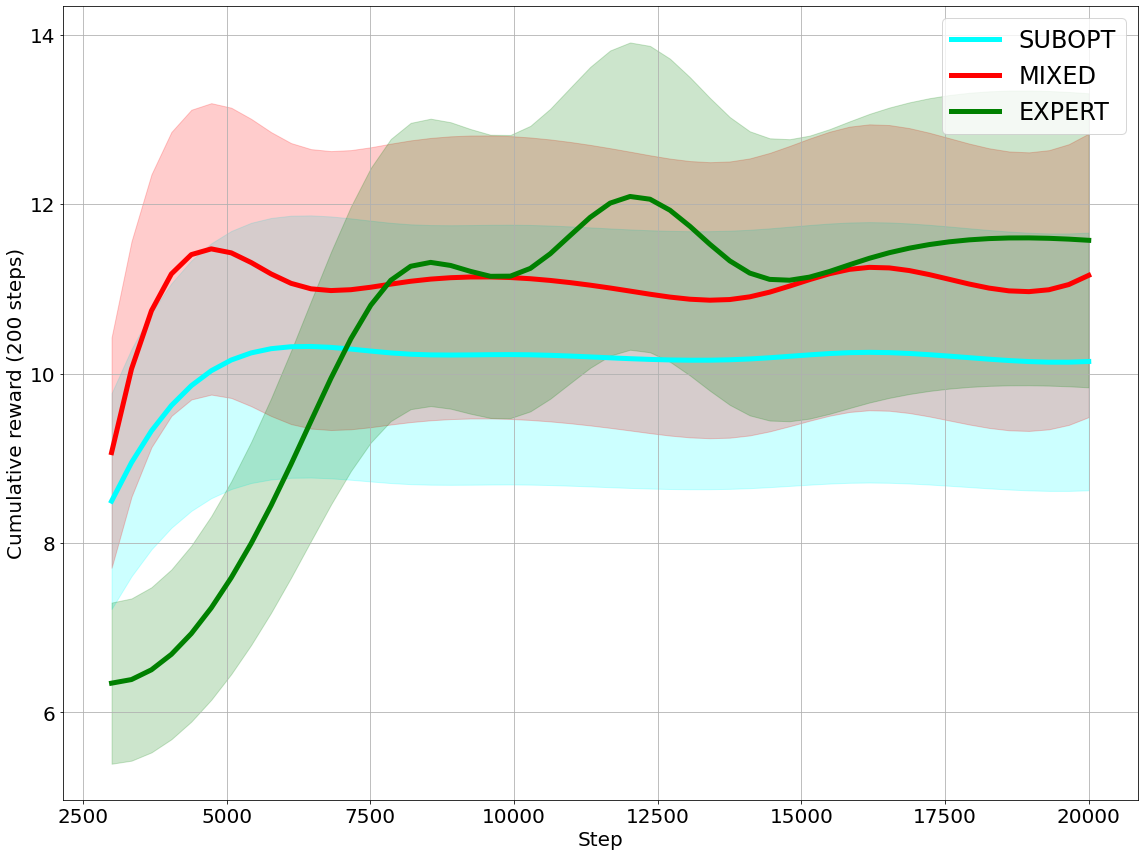}
    \caption{Offline training results with sub-optimal datasets.}
    \label{fig:subopt}
\end{figure}

The outcomes of these training exercises on the sub-optimal datasets are illustrated in Figure \ref{fig:subopt}. From these results, we observe that the performance achieved with the sub-optimal datasets notably surpasses that of the corresponding BPs, yet falls marginally short of the expert-level performance. This finding underscores the capability of offline RL to extract valuable learning even from less-than-ideal data.

Furthermore, an interesting development emerged when we amalgamate the load-oriented and delay-oriented baseline datasets for training. This blend of datasets, encompassing a broader spectrum of network scenarios and challenges, enabled the offline RL agent to approach, and in some cases match, the performance level of the expert-level dataset. This enhancement in performance indicates that \textit{diversity} and \textit{comprehensiveness} in the training dataset play a crucial role in the efficacy of offline RL. It suggests that by judiciously combining datasets with varied characteristics, we can equip the offline RL model with a richer learning experience, thus enabling it to develop more robust and effective strategies that are closer to those derived from optimal conditions. This approach might be particularly beneficial in practical scenarios, where access to expert-level data is limited, and reliance on diverse sub-optimal data sources is more realistic.

\begin{figure*}
\centering
\includegraphics[width = .85\textwidth]{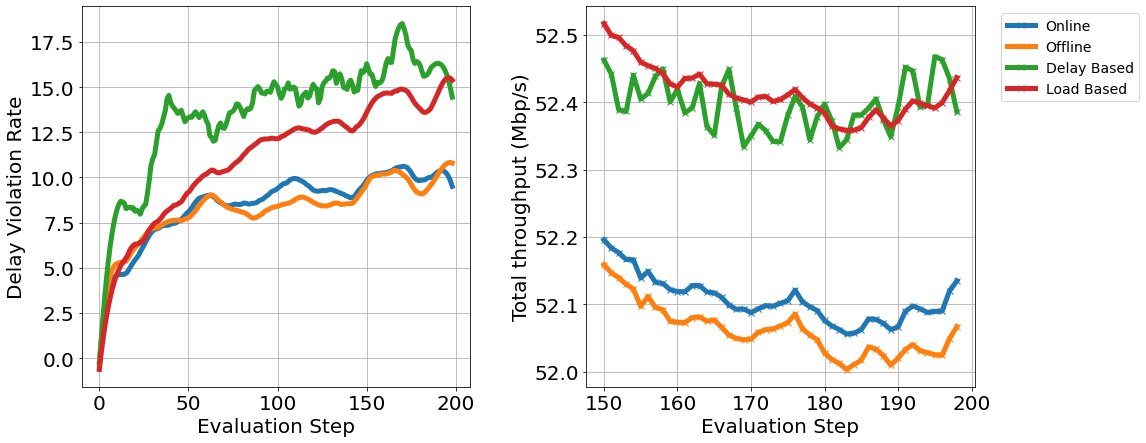}
\caption{Delay violation rate (average of two slices) and total throughput (sum of two slices). The results are averaged over 20 different environments as discussed in Section \ref{subsec:subopt}.}
\label{fig:subopt_behavior}
\end{figure*}

\mypara{Behavior Understanding and Performance Analysis:} Beyond the encouraging cumulative reward outcomes observed during the training phase, it is crucial to delve deeper into the underlying factors contributing to the superior performance of the RL algorithms over the baselines. To this end, we conduct a thorough analysis across 20 distinct environments, encompassing five different user distributions, each evaluated with four unique random seeds. This comprehensive test allows us to assess the robustness and adaptability of the policies in various scenarios. The results, focusing on key metrics like throughput and delay violation rate, are presented in Figure \ref{fig:subopt_behavior}. As illustrated in the figure, the RL algorithms demonstrate the ability to sustain the throughput performance (0.3 Mbp/s drop total throughput wise) while simultaneously achieving a significant reduction in delay violation rates, with a relative improvement of approximately $50\%$. This impressive feat is attributed to the adoption of a more conservative resource allocation strategy compared with the load or delay-aware baselines. Unlike these baseline methods, which may resort to drastic adjustments leading to increased delay violations or throughput drops, the RL algorithm implements a more balanced approach. This strategy effectively mitigates the risk of extreme resource allocation decisions that could be detrimental to overall system performance. The results clearly showcase the proficiency of RL in not only maintaining service quality but also significantly improving network reliability and user experience by reducing delay violations, a critical aspect in RAN slicing environments.

% \congc{Fig.~\ref{fig:subopt_behavior} is not presented well. It is too skewed and not easy to read. I suggest make both sub-figures square. The same issue happens with Fig.~\ref{fig:cross_sla} as well.}

\subsection{Adaptation to Different SLA Requirements}
\label{subsec:adapt_sla}

In the preceding section, we have established that offline RL can surpass online RL in scenarios with consistent SLA requirements, demonstrating the advantage of offline RL in leveraging mixed sub-optimal datasets. In practical RAN slicing systems, however, heterogeneity is not only a result of different BPs but also from distinct SLA requirements. In this context, we explore how offline RL adapts to varying SLAs within a RAN slicing system.

Specifically, we adjust the SLA requirements for Slice 1 by reducing the delay violation threshold from 100 ms to 50 ms and further down to 30 ms. Our goal is to investigate if an offline RL policy can effectively utilize datasets collected under different SLAs to adapt to new SLA conditions that have not been seen before.

\begin{figure*}[hpbt]
\centering
\includegraphics[width = .85\textwidth]{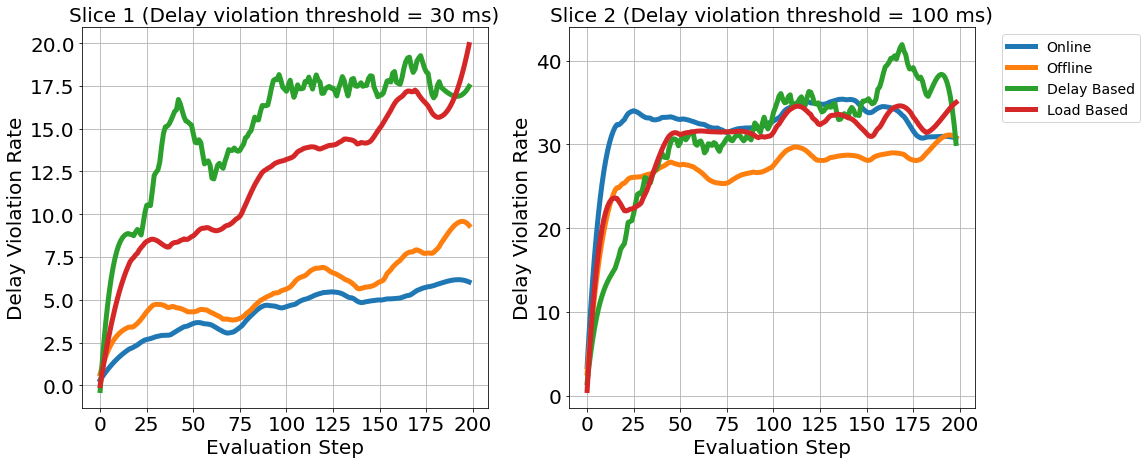}
\caption{Delay violation rate comparison across slices with different SLA requirements.}
\label{fig:cross_sla}
\end{figure*}

For this experiment, we train an offline RL policy using data collected at delay violation thresholds of 100 ms and 50 ms. We then test this policy in an environment where Slice 1 has a delay violation threshold of 30 ms. The performance of this offline RL policy is compared against an online RL policy: one trained in an environment with exact 30 ms threshold. The comparative performance is illustrated in Figure \ref{fig:cross_sla}. The result reveals that both RL methods successfully adapt to the new SLA requirement. Notably, both methods slightly compromise the performance on Slice 2 to mitigate substantial delay violations resulting from the altered SLA on Slice 1. In Figure \ref{fig:cross_sla}, a trade-off of less than $1\%$ in the mean delay violation rate for Slice 2 leads to a reduction of over $10\%$ in delay violations for Slice 1, translating to a relative improvement of over $100\% $ compared with the best-performing baseline. It is more exciting that this is accomplished while the offline policy has never seen any data collected from 30 ms SLA environment. Despite having no prior exposure to this specific SLA requirement, it manages to achieve a performance level comparable to that of the online RL agent, which necessitates tens of thousands of steps of environmental interaction.

\subsection{Variation on Reward Functions}
\label{subsec:vary_reward}

One unique advantage of offline RL in the context of RAN slicing lies in its ability to adjust the reward function based on existing datasets, thereby enabling tailored policy behavior. For instance, in our initial setup, we set the parameters $\alpha = 4$ and $\delta = 1$ to emphasize the impact of delay violation. To shift the focus towards throughput, we adjust the reward parameters to $\alpha = 0.5$ and $\delta = 0.5$. Subsequently, to underscore resource usage, we modify them to $\alpha = 1$ and $\delta = 4$. These modifications in the reward function are expected to significantly influence the behavior of the offline RL policy, as demonstrated in Table \ref{tab:reward}.

\begin{table*}[hpbt]
\caption{Performance Comparison for Different Reward-oriented Offline RL Policies}
\label{tab:reward}
\centering
\begin{tabular}{c|c|c|c}
\hline
 &  Mean delay violation rate & Total throughput (Mbp/s) & Mean resource usage \\
\hline
$CQL_{\text{delay}}$ & $\mathbf{6.5 \pm 3.5}$ & $52.48 \pm 13.65$ & $ 49.15 $\\
$CQL_{\text{throughput}}$ & $9.1 \pm 4.4$ & $\mathbf{58.68 \pm 11.23}$ & $ 49.35 $\\
$CQL_{\text{resource}}$ & $7.3 \pm 4.1$ & $51.44 \pm 12.68$ & $ \mathbf{48.89} $ \\
\hline
\end{tabular}
\end{table*}

The result in Table \ref{tab:reward} reveals that adapting the reward parameters indeed changes the policy behavior. The $CQL_{\text{throughput}}$ policy, with a reduced emphasis on delay violation, yields the highest total throughput but at the cost of increased delay violations. Meanwhile, $CQL_{\text{delay}}$, with its focus on minimizing delays, has the lowest mean delay violation rate. Interestingly, the $CQL_{\text{resource}}$ policy, aimed at optimizing resource usage, does not demonstrate a marked improvement in resource efficiency. This is particularly notable in our tested limited soft-slicing system, where shared resources inherently limit the scope for significant optimization. This finding suggests that the system architectural setup -- soft versus hard slicing -- plays a critical role in determining the efficacy of different reward-oriented policies. In hard slicing environments, where resources are exclusively allocated, the potential for a resource-oriented policy to improve utilization may become more pronounced.

\section{Discussion}
\label{sec:dis}

We discuss several key insights gleaned from the experiments with offline RL in the context of RAN slicing.

\mypara{Efficiency Gains through Reduced Interactions:} Our use of the \texttt{netgymenv} simulator, a near-real-world tool developed by Intel Labs that is based on NS-3 \cite{riley2010ns}, offers practical and credible experimental outcomes. However, it also highlights the increased cost associated with environmental interactions in more sophisticated simulators. In contrast to lightweight simulators where interactions are nearly cost-free and instantaneous, each interaction with \texttt{netgymenv} incurs a substantial delay of approximately 200 ms, encompassing both simulation and communication overheads (this could be optimized with improved networking solutions). In comparison, a single neural network update step takes about 100 ms. Consequently, offline RL, in addition to reducing resource interactions and circumventing sub-optimal exploratory steps inherent in online RL, offers significant time savings. In our experiments, offline RL achieves a time reduction of at least $50\%$, potentially extending up to $67\%$.

\mypara{The Crucial Role of State Normalization:} As outlined in Section \ref{sec:prob}, the initial state definitions utilize raw data metrics including throughput, delay violation rate, and delay. However, we have observed that directly using these raw values could lead to instability in the RL model convergence, sometimes resulting in the policies getting trapped at sub-optimal performance levels. To address this issue, we have found that normalizing the state values is extremely beneficial. By scaling all state values to a normalized range of $[0,1]$, we significantly enhance the stability of the training process. Therefore, we advocate for the implementation of state normalization in future experiments involving RL, as it markedly improves training stability and the overall effectiveness of the RL model.

% \section{Conclusion and Future Directions}
\section{Conclusion}
\label{sec:con}

This work established that offline RL can effectively extract (near)-optimal policies from sub-optimal datasets, highlighting a key advantage over online RL, especially in the context of real-world wireless systems. This is attributed to the capability of offline RL to operate without the need for costly environmental interactions. Additionally, we observed that offline RL is adept at adapting to varying Service Level Agreement (SLA) requirements, demonstrating promising transferability even to previously unseen SLA scenarios. Another important finding was the flexibility of offline RL in policy adjustment. By altering reward functions offline, it is feasible to train multiple policies with distinct objectives. We believe this work is helpful to enhance the current workflow of applying RL to RRM in RAN slicing systems.

% As aforementioned, the main focus of this paper was to shed light of using offline RL in a RAN-slicing system. We found that there are still a lot open problems remaining to be solved in this direction. For example, in our work, we can vary the user density and the SLA requirements for each slice, but not able to vary the traffic patterns. In is really interesting to see whether the offline RL still have the adaptation ability with slices having different traffic flows. One other thing we found critical is the ability to scale up the system, whether it is scaling up the number of the cells or the number of the slices. For a practical system, this feature is almost as important as adaptability. Furthermore, currently we have different slices to share the same reward design, while in practical, different slices could have different SLA requirement based on the service, understanding the capability of customized reward design for each slice is also an interesting topic for future investigation.

\bibliographystyle{IEEEtran}
\bibliography{bibtex}

\end{document}